\documentclass{aa} % LaTeX A&A  Standard Fonts
\usepackage{epsfig}

\begin{document}

%   \thesaurus{06         % A&A Section 06: Formation, structure, evolution of
%                         %                 stars
%           (08.02.4, % Stars: binaries: spectroscopic
%           08.09.2;  % Stars: individual: UX Ari
%           08.12.1; % Stars: late-type
%           08.01.2; % Stars: activity
%           03.20.7; % Techniques: radial velocities
%             )}
%
   \title{The radial velocities of the RS~CVn star UX~Ari
          \thanks{
          based on observations collected at the Nordic Optical
          Telescope (NOT), European Northern Observatory, La Palma, Spain,
          and at the McDonald and Kitt Peak Observatories, USA}
          \fnmsep\thanks{Table~2 is  only available in electronic form
                     at the CDS via anonymous ftp to cdsarc.u-strasbg.fr
                     (130.79.128.5) or via
                     http://cdsweb.u-strasbg.fr/Abstract.html}
         }

   \subtitle{A triple system with a binary on the same line of sight}

   \author{R.~Duemmler\inst{1}
   \and V.~Aarum\inst{2}
   }

   \institute{ Astronomy Division,% University of Oulu, Linnanmaa,
               P.O.\ Box 3000, FIN--90014 University of Oulu, Finland
             \and Institute of Theoretical Astrophysics, University of Oslo,
                 P.O.\ Box 1029 Blindern, N--0315 Oslo, Norway
             }
   \offprints{R.\ Duemmler\\
     \email{Rudolf.Duemmler@Oulu.Fi}}
   \date{Received date; accepted date}

\abstract{
UX~Ari\ belongs to the class of very active RS~CVn stars and has recently
been the target of surface (Doppler) imaging. Although this technique needs a
quite accurate determination of the orbit (in order to have the correct period
for phasing and the correct Doppler shift correction of the line profiles) we
found only one, quite old orbit solution, which has subsequently been used by
everyone.\\
We used published radial velocities (RVs), supplemented by a large number
(124) of our own recent, high-accuracy RVs of both the primary (K0\,IV) and
the secondary (G5\,V) to improve the orbit of UX~Ari. In addition to the
improved set of parameters, we found that the $\gamma$ velocity of the system
is systematically changing over time. It seems that UX~Ari\ is a triple system.
Actually, a third star is weakly present in the spectrum.
While its RV is also changing, it is not a member of the system, but happens
to be on the same line of sight.\\
% Thus, we have on the same
%line of sight a spectroscopic triple system with a spectroscopic binary in
%the background.\\
Finally, conclusions about the  physical parameters of the objects from
the orbits are presented.
\keywords{Stars: individual: UX~Ari -- Stars: binaries: spectroscopic --
        Stars: late-type -- Stars: activity -- Techniques: radial velocities}
} %end of abstract

   \maketitle
%   \markboth{}{}

\section{Introduction}
\label{S-intro}

UX~Ari\ (HD 21242; K0\,IV + G5\,V) is a short-period ($P\approx 6\fd4$),
double-lined spectroscopic binary. It belongs to the class of RS~CVn stars,
i.e.\ at least the cool primary shows signs of activity. Thus, it is listed
in the catalogue by Strassmeier et al.\ (\cite{SHFS93}),
where more information on the system may be found (UX~Ari=CABS\,28);
it was also put in our long term programme of
surface imaging of active stars (for results %of this programme
see e.g.\ Berdyugina et al.\ \cite{B98}).

Surface imaging needs good orbital parameters. The spectral lines, whose
distortions are followed through the rotational phases\footnote{given the
evolved state of one of the components and the short period, it is reasonable
to assume synchronized rotation, i.e.\ $P_{\rm rot}=P_{\rm orb}$} need to have
the Doppler shift due to the orbital motion removed; while experience shows
that small residual shifts do not change the main surface structures, the
extra noise in the data might lead to a lower quality of the map.
Normally, the cross-correlation technique can be used to align spectra and
remove radial velocity shifts without knowledge of any orbital parameters.
However, cross-correlation relies on the assumption that the spectral features
in the programme spectrum and the template are identical, which they are not
in the case of active stars due to the distortions caused by spots. This
leads to systematic radial velocity errors. It is hoped that if one compiles
a large set of radial velocities spanning a long time, the constraints of
orbital motion and the limited lifetime of spots allow the determination
of a good set of orbital parameters despite the fact that individual radial
velocities obtained in a short time span are systematically shifted.
In order to calculate
the rotational phases, a good value for the period needs to be known. This is
especially important if maps obtained during several seasons are to be
compared: an incorrect period and conjunction time lead to increasing phase
shifts
which mimic motions of the surface structure which are not real and lead to
incorrect interpretation of the long term behaviour of the surface
structures. Thus a (re-)determination of the orbit prior to surface imaging
is strongly recommended.

For UX~Ari, there seems to exist only one orbit determination:\ that
by Carlos \& Popper (\cite{CP71}). Given the age and the low number of
measurements used, we felt it long overdue to compute a new orbit and improve
the orbital parameters as much as possible.
%This is also due to the fact
%that Carlos \& Popper seem to have made two independent single-lined orbit
%fits to their data instead of one double-lined fit.
The time difference
between the first observation given by Carlos \& Popper and our last
observation is more than 42 years, which lets us hope to significantly improve
the period and subsequently all other orbital parameters.

Another finding makes an accurate inspection of the orbit %very
particularly interesting.
Lestrade et al.\ (\cite{LPJ99}) performed high-precision VLBI astrometry of,
among others,
UX~Ari. %and ten other radio-emitting stars.
%They observed UX~Ari at 12 epochs
%between July 1983 and May 1994 and calculated the acceleration %accelerations
%$\gamma_{\alpha}$ and $\gamma_{\delta}$
%of UX~Ari's position in the sky based
%on the 10 epochs when the radio emission of the system was above
%detectability. %The results were
%$\gamma_{\alpha}=-0.000041\pm0.000005\ {\rm s\,yr}^{-2}$ and
%$\gamma_{\delta}=-0.00029\pm0.00007\ {\rm ''yr}^{-2}$.
They used observations at 10 epochs between July 1983 and May 1994 to calculate
the acceleration of UX~Ari's position in the sky.
The resulting
acceleration was much larger than the perspective secular change in proper
motion %(which is $1.2\cdot10^{-7}\ {\rm ''yr}^{-2}$)
and could be caused by a third body. According to Lestrade et al.\
(\cite{LPJ99}),
the orbital period of this system should be many times their
11-year VLBI data span.

\section{Observations and data reductions}
\label{S-obs}

\begin{table}
\caption{\label{T-rsn} The wavelength resolution ($R=\lambda/\Delta\lambda$)
                       and range in signal-to-noise ratios (S/N)
                       at 6400\,\AA\ for each dataset.}
\begin{center}
\begin{tabular}{|l|c|c|}
\hline
Set& $R$  & S/N\\
\hline
S95 & 86\,000 & 270--420\\
S96 & 71\,000 & 280--390\\
M99 & 36\,000 & 180--390\\
K99 & 86\,000 & 90--170\\
M00 & 48\,000 & 140--360\\
\hline
\end{tabular}
\end{center}
\end{table}

The new high-resolution, high signal-to-noise spectra were obtained for
the purpose of surface imaging during five observing runs using three
telescope-instrument combinations:
\begin{itemize}
\item 5 spectra obtained in December 1995 using the high-resolution
\'echelle spectrograph SOFIN (Tuominen \cite{T92}) mounted at the Cassegrain
focus of the  2.56\,m Nordic Optical
Telescope (NOT) on La Palma, Canary Islands, Spain (hereafter denoted S95)
\item 10 spectra obtained in November/December 1996 using NOT/SOFIN (hereafter
denoted S96)
\item 64 spectra obtained in January 1999 using the Sandiford Cassegrain
\'Echelle Spectrograph mounted on the 2.1\,m Otto Struve Telescope at McDonald
Observatory, Texas, USA (hereafter denoted M99)
\item 23 spectra obtained in February 1999 using the Coud\'e CCD Spectrograph
receiving light from the 0.9\,m Coud\'e-Feed Telescope at Kitt Peak National
Observatory, Arizona, USA (hereafter denoted K99)
\item 22 spectra obtained in January 2000 again using the 2.1\,m telescope with
the Sandiford spectrograph at \hbox{McDonald} Observatory
(hereafter denoted M00)
\end{itemize}

Table~\ref{T-rsn} shows the wavelength resolution ($\lambda/\Delta\lambda$) at
6400\,\AA\ and the range of signal-to-noise ratios at 6400\,\AA\ for each
dataset. The spectrograph slit widths were chosen so that one resolution
element (FWHM of the ThAr comparison lines) was
% between 2 and 3 pixels on the CCD\@.
2--3 CCD-pixels.
The spectra in K99 have considerably lower signal-to-noise ratios than
the other datasets due to the smaller telescope.

The spectra in S95 and S96 were reduced using the 3A Software Package
(Ilyin \cite{I96}). It uses two-dimensional dispersion curves (rows
and columns of the CCD; see
e.g.\ Duemmler et al.\ (\cite{DIT97}) for a more thorough description) in
order to calibrate the wavelength scale. The spectra in M99 were reduced
using the 4A Software Package (Ilyin \cite{I00}). The spectra in K99 and
M00 were reduced using IRAF\@. 4A and IRAF use three-dimensional dispersion
curves (rows and columns of the CCD as well as time on the basis of several
comparison spectra) to calibrate the
wavelength scale. For all spectra, telluric lines based on the wavelengths
given by Pierce \& Breckinridge (\cite{PB73}) were
used prior to heliocentric correction to establish the accurate wavelength
zero point, correcting for tiny
geometrical shifts between the comparison and stellar images due to bending of
the spectrograph and the slit-effect, i.e.\ the shifts due to the fact that the
optics is not homogeneously illuminated by the stellar light %.
(Griffin \& Griffin \cite{G73}, Ilyin \cite{I00}).

The projected equatorial rotational velocity $v\sin i$ of the subgiant primary
in the RS~CVn binary was determined from our measurements using a
Fourier-transform technique described by Gray (\cite{G88}; \cite{G92}, Ch.~17).
The result is $v\sin i=39$\,km\,s$^{-1}$. For the secondary, a value of
$v\sin i=7.5$\,km\,s$^{-1}$\ was determined by comparing spun-up standard spectra with
that of the secondary, a value consistent with the one given by
Vogt \& Hatzes (\cite{VH91}).

The radial velocities (RVs) were obtained by cross-correlating the UX~Ari\ spectra
with spectra of RV standard stars. For the primary we used spectra of
$\beta$\,Gem (K0\,IIIb, RV $=+(3.3\pm0.1)$\,km\,s$^{-1}$) observed
%at the same time 
in the same run as the
UX~Ari\ spectra, reduced in the same way and artificially spun up to match
$v\sin i$ of the primary.
For the secondary we used the solar
%(G2\,V, RV $=0.0$\,km\,s$^{-1}$)
FTS spectrum (Kurucz et al.\ \cite{KFBT84}),
artificially spun up to match $v\sin i$ of the secondary. The RVs are
weighted averages of the individual RVs measured in several orders.
The measured RVs and their standard deviations are given in
%Tab.~\ref{T-RV}\footnote{Table~\ref{T-RV} is only
Table~2\footnote{Table~2 is only
available in electronic form at the CDS via anonymous ftp to
\hbox{cdsarc.u-strasbg.fr} (130.79.128.5) or via
\hbox{http://cdsweb.u-strasbg.fr/Abstract.html\,.}}.

\setcounter{table}{2} %Tab.2 is commented out, make sure that next table is Tab.3

%In the spectrum of UX~Ari there are also weak lines from a third star present.
%This star
%was first mentioned by McAlister et al.\ (\cite{MHH87}), but it is not known
%whether this star is part of the UX~Ari\ system or a more distant star that just
%happens to lie on the same line of sight. Its spectral classification is also
%not known. Vogt \& Hatzes (1991) used a synthetic G5\,V spectrum to subtract
%the lines of the third star from the composite spectrum, and we used the solar
%FTS spectrum as RV template to measure the radial velocity of this star.

\begin{table}
\caption{\label{T-angsep} Measurements from speckle interferometry and from
       Hipparcos (ESA \cite{ESA97}) of the angular separation $\vartheta$
       between the  RS~CVn system and the third star at different epochs
       (as Besselian year).}
\begin{center}
\begin{tabular}{|l|l|l|}
\hline
Epoch  & $\vartheta$ ($''$) & Reference\\
\hline
1985.8431 & 0.432 & McAlister et al.\ (\cite{MHH87})\\
1991.25   & 0.340 & Hipparcos (ESA \cite{ESA97})\\
1995.9237 & 0.297 & Hartkopf et al.\ (\cite{HMM97})\\
1996.8658 & 0.256 & Hartkopf et al.\ (\cite{HMM00})\\
\hline
\end{tabular}
\end{center}
\end{table}

In the spectra of UX~Ari there are also weak lines from a third star present.
This star was first mentioned %in
by McAlister et al.\ (\cite{MHH87}) %where
when they
measured the angular separation between the RS~CVn binary and the third star.
This and other measurements of the angular separation are given in
Table~\ref{T-angsep}. From Table~\ref{T-angsep} it seems that the angular
separation
%between the RS~CVn binary and the third star
has decreased by almost
$0\farcs2$ from 1985.8 to 1996.9. It is not known whether the third star is
part of
the UX~Ari system or a %more distant
%closer or more distant
star that just happens to lie on the same
line of sight. Its spectral classification is also not known, although
Fabricius \& Makarov (\cite{FM00}) determined its B and V magnitudes based on
Hipparcos (ESA \cite{ESA97}) data. Their B and V magnitudes yield
${\rm B}-{\rm V}=1.19\pm0.06$ for the third star. This in turn yields a spectral
type of K5 if the star is unreddened and on the main sequence (Gray \cite{G92}).
Vogt \& Hatzes (\cite{VH91}) successfully used %successfully
a synthetic G5\,V spectrum to
subtract the lines of the third star from the composite spectrum,
and we used the solar FTS spectrum as RV template to measure the radial
velocities of this star.

Before RV measurements of the third star could be carried out, however, the
spectral flux
contributions from the primary and the secondary had to be removed from the
composite spectrum. Otherwise, the weak lines of the third star are too
strongly influenced by the stronger lines of the other two components.

The spectral flux contribution from the primary (secondary) was removed using
an observed spectrum of a single, inactive star of the same spectral
classification as the primary (secondary). We used HD~71952 (K0\,IV,
V$=6\fm25$) and HD~84453 (K0\,IV, V$=6\fm83$) for the primary and HD~23565
(G5\,V, V$=7\fm70$), HD~51419 (G5\,V, V$=6\fm94$) and HD~71148 (G5\,V,
V$=6\fm30$) for the secondary. The single star was observed
in the same run
%at the same time
as
UX~Ari\ %using the same telescope-instrument combination
and reduced in the same
way. Its spectrum was artificially spun up to match $v\sin i$ of the
primary (secondary), scaled to match the relative continuum flux contribution
of the primary (secondary) and shifted by cross-correlation to the position of
the primary (secondary) in the composite spectrum. Finally, the spun-up, scaled
and shifted spectrum of the single star was subtracted from the composite
spectrum. The relative continuum flux contribution for each component was
determined using the residual line strength in the composite spectrum.

The spectrum separation technique itself, as well as what applying it to our
UX~Ari observations can teach us about the three components in the UX~Ari
spectrum, will be described by Aarum \& Engvold (in preparation). The results
of applying the surface (Doppler) imaging technique to our UX~Ari observations
(and thus the details of the line profiles) will be described by Aarum
et al.\ (in preparation).

\section{The radial velocity curves of UX~Ari}
\label{S-RVshort}

\subsection{The data from the literature}
\label{SS-lit}

A significant improvement of all orbital parameters depends strongly on the
value of the period, which in turn is more sensitive to the overall time span
covered by the measurements than to their quality. Thus, we supplement our new
radial velocities with data from the literature to increase the time span.

There are not %very
many RVs of UX~Ari\ to be found. The oldest dataset is due
to Carlos \& Popper (\cite{CP71}). It contains a few RVs obtained in the
1950s, but mostly data from 1967 to 1970; the total number of RV pairs is
28. The second big dataset is due to Duquennoy et al.\ (\cite{DMH91}),
containing 36 timepoints with RVs (however often only for one of the
two stars) obtained mostly in 1977; a few RVs are %taken in the years
measured in
1985--1988. Duquennoy et al.\ (\cite{DMH91}) give only the RVs;
they do not determine or improve the orbital parameters. Another
dataset is given by Heintz (\cite{H81}). There are only 3 RVs for each component
given, and, when compared to a preliminary orbit, they have considerable
scatter. This would give them such a low weight in the combined dataset that,
together with their small number and the fact that their observing times
overlap with the dataset of Duquennoy et al.\ (\cite{DMH91}),
we decided not to use them at all.

\subsection{The weights}
\label{SS-weights}

The optimal weights in a least squares fit are the inverse variances of the
individual measurements.
For all datasets, individual error estimates for the RVs are known, except
for the set given by Carlos \& Popper (\cite{CP71}). Thus, we decided to use
the inverse variance as the weight, and determine an estimate for this for
the Carlos \& Popper set.

An independent orbital fit of a double-lined binary RV curve to the data
of Carlos \& Popper (\cite{CP71}) alone was performed, using their relative
weights\footnote{except for the 3 very old measurements, which had weights 0
in their final fit, but obtained 0.7 here; finally, since it
improved the fit significantly, all data with RV-differences
$< 40$\,km\,s$^{-1}$\ between primary and secondary obtained \hbox{weights 0}}.
The resulting orbital parameters are close to those given by Carlos \& Popper.
Standard deviations for measurements having unit weight for the primary
(1.6\,km\,s$^{-1}$) and the secondary (1.9\,km\,s$^{-1}$) were obtained and used
to calculate for each measurement an error by combining them with the
relative weight.
% given by Carlos \& Popper (\cite{CP71}).
These are the errors given in Table~2 %Tab.~\ref{T-RV}
for the measurements of Carlos \& Popper (\cite{CP71}) and used to calculate
the weights as the inverse variances.

All other measurements obtained their weights as the inverse variances based
on the errors as published. Preliminary orbit fits, however,
indicated that a considerable improvement is achieved when ignoring all
measurements obtained from blended lines, i.e.\ with RV-differences
$< 40$\,km\,s$^{-1}$\ between primary and secondary. Therefore, we decided to
give
all those measurements %zero weight
weight 0
in the following; they are indicated by
negative errors in Table~2.%Tab.~\ref{T-RV}.

During the first orbital fits it turned out that $\chi^2$ for the primary
is significantly larger than $\chi^2$ for the secondary, although the smaller
accuracy of the RVs due to the much broader lines is already reflected by the
much larger RV-errors. This could be due to the systematic deviations  of the
RVs caused by the fine-structure of the line profiles because of the spot
activity. We therefore multiplied all weights of the primary by an additional
factor 0.65 to equalize $\chi^2$ for the primary and the secondary.

\subsection{The first fit results: $\gamma$ is changing}
\label{SS-Fit1}

\begin{table*}[tb]
\caption{\label{Fit1} %FIT037 in UXAri/NEWRV/uxari.par for e=0, FIT038 for e>0
        Two orbital solutions using the data and errors given in
        Table~2; %Tab.~\ref{T-RV};
        the weights of the primary have been additionally
        reduced by a factor 0.65. Fit 1: circular, 7 different $\gamma$ velocities
        allowed; the index of $\gamma$ identifies the dataset, where CP stands
        for Carlos \& Popper (\cite{CP71}) and DMH for
        Duquennoy et al.\ (\cite{DMH91}).
%$\gamma_{\rm CP}$ for the data by Carlos \& Popper (\cite{CP71}),
%        $\gamma_{\rm DMH}$ for the data by Duquennoy et al.\ (\cite{DMH91}),
%        $\gamma_{\rm S95}$ for the data obtained with SOFIN in 1995,
%        $\gamma_{\rm S95}$ for the SOFIN-data of 1995,
%        $\gamma_{\rm S96}$ for the data obtained with SOFIN in 1996,
%        $\gamma_{\rm M99}$ for the McDonald-data of 1999,
%        $\gamma_{\rm K99}$ for the data obtained at Kitt Peak in 1999,
%        $\gamma_{\rm M00}$ for the data obtained at McDonald Observatory in 2000.
        Fit 2: like Fit 1, but allowing for
        $e > 0$. C\&P: parameters of Carlos \& Popper (\cite{CP71}).
      }
\begin{center}
\begin{tabular}{|l|r@{$\,\pm\,$}l|r@{$\,\pm\,$}l|r@{$\,\pm\,$}l|}
\hline
parameter                         & \multicolumn{2}{c|}{Fit 1}        & \multicolumn{2}{c|}{Fit 2}       & \multicolumn{2}{c|}{C\&P\,$^1$} \\[0.1cm]
\hline
$P_{\rm obs}$ (days)              & 6.4378553                      & 0.0000046                    & 6.4378564               & 0.0000046     & 6.43791              & 0.00008  \\
$K_1$ (km\,s$^{-1}$)              & 57.88                          & 0.17                         & 57.93                   & 0.17                     & 59.4                 & 0.6      \\
$K_2$ (km\,s$^{-1}$)              & 66.978                          & 0.033                         & 66.971                   & 0.033                     & 66.7                 & 0.8      \\
$e$                               & \multicolumn{2}{c|}{0.0 (fixed)} & 0.0018           & 0.0007\,$^2$ & \multicolumn{2}{c|}{0.0 (fixed)}\\
$\omega$ (deg)                    & \multicolumn{2}{c|}{---}         & 31.7                 & 20.0                   & \multicolumn{2}{c|}{---}        \\
$T_0$ (HJD)\,$^3$                  & 2450642.00075                  & 0.00077                        & 2450642.57            & 0.36                   & \multicolumn{2}{c|}{---} \\
$T_{\rm conj}$ (HJD)\,$^4$         & 2450640.39129                 & 0.00077                        & 2450640.39            & 0.51                   & 2450640.44          & 0.13\,$^5$     \\
$a_1\,\sin\,i$ ($R_{\sun}$)       & 7.362                          & 0.021                          & 7.368                    & 0.021                 & 7.55                 & 0.13\,$^6$ \\
$a_2\,\sin\,i$ ($R_{\sun}$)       & 8.5192                          & 0.0042                          & 8.5183                    & 0.0042                & 8.50                 & 0.13\,$^6$ \\
$m_1\,\sin^3\,i$ ($M_{\sun}$)     & 0.6964                         & 0.0066                         & 0.6968                   & 0.0066                    & 0.71                 & 0.01       \\
$m_2\,\sin^3\,i$ ($M_{\sun}$)     & 0.6018                         & 0.0055                         & 0.6027                   & 0.0054                    & 0.63                 & 0.01       \\
$\gamma_{\rm CP}$  (km\,s$^{-1}$)  & 26.46                        & 0.75                          & 26.47                   & 0.74                     & 26.5                 & 0.6        \\
$\gamma_{\rm DMH}$  (km\,s$^{-1}$)  & 27.30                       & 0.23                          & 27.29                   & 0.23                     & \multicolumn{2}{c|}{---} \\
$\gamma_{\rm S95}$  (km\,s$^{-1}$) & 28.806                      & 0.047                          & 28.911                   & 0.061               & \multicolumn{2}{c|}{---} \\
$\gamma_{\rm S96}$  (km\,s$^{-1}$) & 29.273                      & 0.049                          & 29.309                   & 0.053               & \multicolumn{2}{c|}{---} \\
$\gamma_{\rm M99}$  (km\,s$^{-1}$) & 28.043                      & 0.067                          & 28.094                   & 0.070                 & \multicolumn{2}{c|}{---} \\
$\gamma_{\rm K99}$  (km\,s$^{-1}$) & 27.898                      & 0.066                          & 27.945                   & 0.071                & \multicolumn{2}{c|}{---} \\
$\gamma_{\rm M00}$  (km\,s$^{-1}$) & 25.905                      & 0.058                          & 25.921                   & 0.061                 & \multicolumn{2}{c|}{---} \\
$\sigma$ (km\,s$^{-1}$)\,$^7$         & \multicolumn{2}{c|}{1.82,\ 0.25}       & \multicolumn{2}{c|}{1.79,\ 0.25}         & \multicolumn{2}{c|}{---}\\[0.1cm]
\hline
\end{tabular}
\end{center}
{\footnotesize
\begin{itemize}
\item[$^1$] Note, that Carlos \& Popper (\cite{CP71}) give mean errors, which
           have been converted to standard deviations here for consistency.
\item[$^2$] A Lucy-Sweeney $F$-test
           (Lucy \& Sweeney \cite{LS71}, Lucy \cite{L89})
           gives a 97.7\% significance for this eccentricity.
\item[$^3$] For $e=0$, HJD of maximum RV of the
           primary, for $e>0$ that of periastron passage.
\item[$^4$] HJD of the conjugation with the secondary (hotter star) in the back.
\item[$^5$] computed from their value (``earlier star'') and their period with
           error propagation.
\item[$^6$] computed from their $a\,\sin\,i$ and their mass-ratio, retaining
           their error for $a\,\sin\,i$.
\item[$^7$] standard deviation of a single RV of mean weight, separately for
           primary and secondary, respectively.
\end{itemize}
} % end of footnotesize
\end{table*}

For each dataset, i.e. Carlos \& Popper (\cite{CP71}),
Duquennoy et al.\ (\cite{DMH91}), and each of the seasons of our own
observations, an individual orbital fit has been done. By this we established:
\begin{itemize}
\item that there is no significant (as compared to the formal fit errors)
     change of any orbital parameter
\item {\it except} the velocity $\gamma$ of the centre of mass of the system,
     which is different for each dataset. %, sometimes by many $\sigma$.
\end{itemize}
The following fit, combining all datasets, thus requires that all orbital
parameters are the same, but allows each dataset to obtain its own velocity
zero-point, i.e.\ its own $\gamma$ velocity.
The results of this fit are given in Table~\ref{Fit1} and compared to the
orbital parameters given by Carlos \& Popper (\cite{CP71}).
Here, and throughout
the paper, errors are formal fit errors, obtained from the curvature of the
$\chi^2$ hypersurface or from error progression, and are likely
underestimates of the true errors.

The results can be summarized as follows:
\begin{itemize}
\item Most orbital parameters given by Carlos \& Popper (\cite{CP71})
     have been confirmed.
\item Due to the much larger database and the %larger
     higher
     accuracy of the later
     data, the accuracy of the orbital parameters has considerably increased.
     The period, in particular, is shorter than that given by
     Carlos \& Popper, while still being consistent with it because of their
     larger error.
\item Despite the %large
     high
     significance given to the eccentricity, we do not
     believe that the orbit is really elliptical. Firstly, the eccentricity is
     very small; secondly, its error is relatively large, so that the
     eccentricity is just a 2.5$\sigma$ detection. Thirdly, as will be seen
     later, there are systematic effects in the radial velocity curve which
     might mimic the effects of a (small) non-zero eccentricity. Finally,
     the deviations from the circular orbit are so negligible that all other
     orbital parameters are indistinguishable from those obtained for the
     circular orbit; thus, no harm is done by neglecting $e$ in the following
     even if it should be real.
\item The values of the velocity of the centre of mass $\gamma$
      of the system are clearly inconsistent with each other and need
      further consideration.
\end{itemize}

While it is not uncommon to have different velocity zero-points from different
instruments, especially when old data are involved, the situation here is
different. Firstly, there are two pairs of datasets (S95, S96 and
M99, M00) which were taken by the same instruments, however one year apart.
The $\gamma$ velocities are inconsistent within the groups which only becomes
apparent thanks to the high quality and number of the radial velocities leading
to really small errors in $\gamma$. Secondly, all new data are reduced in the
same way. This means in particular that the wavelength scale for each spectrum
is adjusted (prior to heliocentric correction) to the same wavelength system
established by a large number of
telluric atmospheric lines based on the wavelengths given by Pierce \&
Breckinridge (\cite{PB73}). This technique should remove all effects caused by
the slit-effect
%(the position of the star not being in the centre of the
%slit, causing a non-homogeneous illumination of the optics; Griffin \&
%Griffin \cite{G73},
%Ilyin \cite{I00})
and temporal effects like those caused by the change of ambient temperature
and pressure and (for the Cassegrain-spectrographs) bending of the
spectrograph due to motion of the telescope.
Furthermore, all RVs were measured using the same star as %a
template
(the IAU RV-standard $\beta$~Gem for the primary
and the solar FTS-spectrum for the secondary;
while $\beta$~Gem has been observed during the same runs with the same
instruments as UX~Ari, and reduced in the same way, the solar FTS-spectrum
used is always the same).
Additionally, the paper by Duquennoy et al.\ (\cite{DMH91}) states that their
radial velocities are on the IAU faint (m$_{\rm V}\geq4.3$) standard system;
yet the difference between $\gamma_{\rm DMH}$ and $\gamma_{\rm S96}$ is
8.4$\sigma$. On the other hand, the difference between
$\gamma_{\rm M99}$ and $\gamma_{\rm K99}$, which were obtained with
different instruments a month
apart, is only 1.5$\sigma$. The difference between $\gamma_{\rm S96}$ and
$\gamma_{\rm M00}$ is 44$\sigma$! Finally, it seems that we have, at least for
the new data, a systematic behaviour of $\gamma$: it seems to increase from
1995 to 1996, and from then on it systematically decreases with time.

For these reasons, we believe that the variation in $\gamma$ is genuine and
not caused by any instrumental effect. The short-period RS~CVn system
UX~Ari\ is obviously in an accelerated motion, most likely around the centre of
mass with a third star.

\subsection{The final fit of the inner orbit}
\label{SS-Fitin}

\begin{table}[tb]
\caption{\label{T-Fitin} %FIT035 in UXAri/NEWRV/uxari.par
        The final orbital solution ($e=0$) using the data and errors given in
        Table~2 %Tab.~\ref{T-RV}
        split into 17 datasets; the weights are as before.
        The period used is always the period in the restframe of the system.
        All parameters are the same for all datasets, except $\gamma$. The
        weighted average HJD for each dataset (--2400000)
        is given together with the fitted $\gamma$.
      }
\begin{center}
\begin{tabular}{|l|r@{$\,\pm\,$}ll|}
\hline
parameter                         &\multicolumn{3}{c|}{ }\\[0.1cm]
\hline
$P_{\rm rest}$ (days)              & 6.4372703                      & 0.0000069 &\\
$K_1$ (km\,s$^{-1}$)              & 57.86                          & 0.17     &\\
$K_2$ (km\,s$^{-1}$)              & 66.980                          & 0.036   &\\
$T_0$ (HJD)                       & 2450642.00204                  & 0.00081  &\\
$T_{\rm conj}$ (HJD)\,$^1$         & 2450640.39272                 & 0.00081   &\\
$a_1\,\sin\,i$ ($R_{\sun}$)       & 7.358                          & 0.022    &\\
$a_2\,\sin\,i$ ($R_{\sun}$)       & 8.5186                          & 0.0045  &\\
$m_1\,\sin^3\,i$ ($M_{\sun}$)     & 0.6962                         & 0.0069   &\\
$m_2\,\sin^3\,i$ ($M_{\sun}$)     & 0.6013                         & 0.0056   &\\
$\gamma_{\rm CP_1}$  (km\,s$^{-1}$)  & 27.6                        & 4.0                          & 34785.6460\\
$\gamma_{\rm CP_2}$  (km\,s$^{-1}$)  & 28.8                        & 1.9                          & 35001.9260\\
$\gamma_{\rm CP_3}$  (km\,s$^{-1}$)  & 27.61                        & 0.97                        & 39813.4043\\
$\gamma_{\rm CP_4}$  (km\,s$^{-1}$)  & 26.4                        & 2.9                          & 39926.6130\\
$\gamma_{\rm CP_5}$  (km\,s$^{-1}$)  & 28.3                        & 1.1                          & 40125.7212\\
$\gamma_{\rm CP_6}$  (km\,s$^{-1}$)  & 28.1                        & 1.4                          & 40519.1555\\
$\gamma_{\rm CP_7}$  (km\,s$^{-1}$)  & 27.5                        & 1.1                          & 40878.2310\\
$\gamma_{\rm DMH_1}$  (km\,s$^{-1}$)  & 27.28                       & 0.23                        & 43440.9205\\
$\gamma_{\rm DMH_2}$  (km\,s$^{-1}$)  & 26.86                       & 0.58                        & 46425.6050\\
$\gamma_{\rm DMH_3}$  (km\,s$^{-1}$)  & 27.67                       & 0.76                        & 46713.0793\\
$\gamma_{\rm DMH_4}$  (km\,s$^{-1}$)  & 26.77                       & 0.94                        & 46845.1837\\
$\gamma_{\rm DMH_5}$  (km\,s$^{-1}$)  & 26.62                       & 0.51                        & 47520.6140\\
$\gamma_{\rm S95}$  (km\,s$^{-1}$) & 28.806                      & 0.049                          & 50055.6248\\
$\gamma_{\rm S96}$  (km\,s$^{-1}$) & 29.277                      & 0.052                          & 50415.4416\\
$\gamma_{\rm M99}$  (km\,s$^{-1}$) & 28.026                      & 0.071                          & 51186.5754\\
$\gamma_{\rm K99}$  (km\,s$^{-1}$) & 27.850                      & 0.067                          & 51215.5638\\
$\gamma_{\rm M00}$  (km\,s$^{-1}$) & 25.877                      & 0.062                          & 51559.1661\\
$\sigma$ (km\,s$^{-1}$)\,$^2$         & \multicolumn{2}{c}{1.93,\ 0.26}       & \\
\hline
\end{tabular}
\end{center}
{\footnotesize
\begin{itemize}
\item[$^1$] HJD of the conjugation with the secondary in the back.
\item[$^2$] standard deviation of a single RV of mean weight, separately for
           primary and secondary, respectively.
\end{itemize}
} % end of footnotesize
\end{table}

If we accept this interpretation, we have to make two changes
in order to obtain the final orbital parameters: First, the old datasets from
the literature cover several years, and we see from the large difference
(as compared to the error) between 1999 and 2000 that $\gamma$ is changing
much over one year; we thus have to subdivide all datasets so that each
subset does not cover more than 1 year. Secondly, if $\gamma$ is changing,
so is the factor between the period in the observer's frame and the period
in the restframe of the system. Given that there is no evidence for any
change of the orbital parameters, except $\gamma$, we can assume that the
period in the restframe of the system is constant, but not the one in the
observer's frame.

\begin{figure}
%\hspace*{-0.3cm}\rotatebox{90}{\resizebox{6.6cm}{!}{\includegraphics{uxari.rvcurve.short.ps}}}
\hspace*{-0.3cm}\rotatebox{90}{\resizebox{6.6cm}{!}{\includegraphics{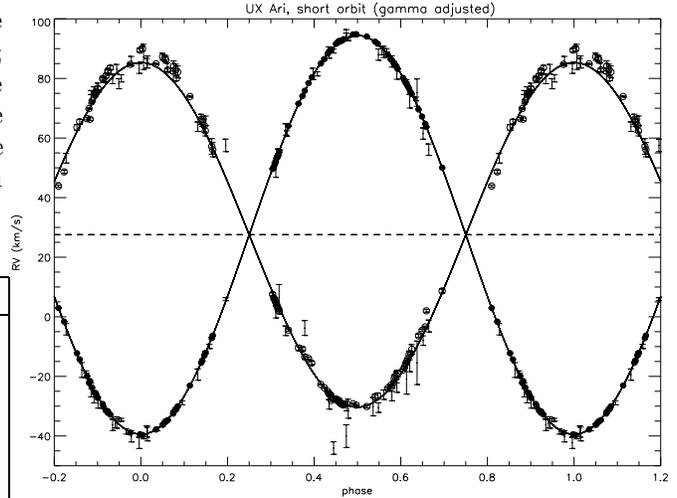}}}
\caption{\label{F-Fitin} Final orbit fit (as given in Table~\ref{T-Fitin}) showing
       as pure errorbars the measurements from the literature; our new
       measurements are shown with filled
       circles for the secondary (the errorbars are usually smaller
       than the symbol size) and as open circles for the primary.
       The different $\gamma$ velocities have been shifted to a common value
       ($\gamma_{\rm DMH_1}$).
       }
\end{figure}

The final orbital fit is thus a weighted fit (using the same weights as before)
to 17 subsets, ensuring that all parameters
have the same values for all datasets, except $\gamma$. Given the small
eccentricity found before, this fit is forced to be circular. The
subsets have been identified in Table~2. %Tab.~\ref{T-RV}.
The resulting orbit is given in Table~\ref{T-Fitin} and shown in
Fig.~\ref{F-Fitin}. Unfortunately, several
of the 17 datasets consist of only 1--3 measurements, so determination of
$\gamma$ is only possible, because no other parameter is determined for these
small datasets alone; nevertheless, for these small datasets and for others
due to large errors in the measurements, the error of $\gamma$ is sometimes
so large that it hides any change of $\gamma$. Due to this and due to the
timing of the older observations, no indication for a changing $\gamma$ could
be detected before our new, accurate datasets.

%We notice that there are only very minor changes in the orbital parameters
%(except $\gamma$).
%We note that the change from $P_{\rm obs}$ to
%$P_{\rm rest}$ leads to a slight deterioration of the fit. This is caused
%by an increase of $\gamma$ for the Carlos \& Popper (\cite{CP71}) datasets by
%roughly 2\,km\,s$^{-1}$; at the same time the O--C becomes
%--2\,km\,s$^{-1}$ for most of these data. We suggest that this is due to the strong correlation
%between $P_{\rm rest}$ and $\gamma$ and a RV zeropoint error in the old data
%of Carlos \& Popper. It does not invalidate our conclusion that the
%short period system is in a wider orbit and its $P_{\rm rest}$ is constant,
%because no such effect is visible in any other dataset, including the ones
%from Duquennoy et al.\ (\cite{DMH91}); the Carlos \& Popper dataset is the
%only one for which we have no information about the RV zeropoint.
%Given the very small influence of these
%measurements on the total fit and the final fit parameters, we did not delete
%these measurements from the overall fit.

In Fig.~\ref{F-Fitin} we see that there is a nearly perfect fit of the RVs of
the secondary. For the primary, however, we notice a large, systematic deviation
near RV-maximum: Most new measurements deviate from the curve towards larger
velocities. No such deviations are seen near the minimum. This is the reason
for the larger $\chi^2$ of the primary as compared to the secondary mentioned
above which motivated us to introduce the extra weight reduction factor of 0.65
for the RVs of the primary. It will be very interesting to see the surface
maps of the primary (Aarum et al., in preparation): %\cite{Aa00}):
when the primary is moving away from us, there should be
a strong surface feature in the facing hemisphere that lets us overestimate
the RV, which is not visible, when the star shows us the other hemisphere.
Also, since the deviation is in nearly all new measurements (which cover more
than 4 years), this feature must be very long-lived. It cannot be a hot spot
in the sub-secondary point, because this would also be visible at RV-minimum.

This also shows that it would not be appropriate
to use cross-correlation techniques
to adjust the spectra for surface imaging: if we would have chosen a spectrum
near RV-maximum as %a
template and adjusted all other spectra to it, we would
have
introduced a significant wavelength shift. It is thus essential that the Doppler
shift corrections prior to surface imaging come from an accurate orbit,
determined over a long time span.

\subsection{The preliminary outer orbit}
\label{SS-Fitout}

\begin{table}[tb]
\caption{\label{T-Fitout} %FIT045(e=0) and FIT039(e>0) in UXAri/NEWRV/uxari.par
        The preliminary orbital solutions to $\gamma(t)$ using the
        $\gamma$ values and average HJDs given in Table~\ref{T-Fitin}. The
        period $P_{\rm out}$ is given in the observer's frame.
      }
\begin{center}
\begin{tabular}{|l|r@{$\,\pm\,$}l|r@{$\,\pm\,$}l|}
\hline
parameter                         &\multicolumn{2}{c|}{circular}
                                  &\multicolumn{2}{c|}{elliptical}\\
\hline
$P_{\rm out}$ (days)              & 3894 & 66  & 7838 & 23\\
$K_{\rm out}$ (km\,s$^{-1}$)      & 2.90  & 0.30 & 2.036 & 0.061 \\
$e$                          & \multicolumn{2}{c|}{0.0 (fixed)} & 0.622 & 0.040\\
$\omega$ (deg.)              & \multicolumn{2}{c|}{---} & 71.7 & 3.6\\
$T_0$ (HJD)                       & 2450495 & 35 & 2451164 & 24  \\
$a_{\rm out}\,\sin\,i$ ($R_{\sun}$)       & 223 & 23 & 247 & 13\\
$f(m)$ ($M_{\sun}$)     & 0.0098 & 0.0031 & 0.00329 & 0.00050  \\
$\gamma_{\rm out}$  (km\,s$^{-1}$)  & 26.53 & 0.22 & 27.23 & 0.12\\
$\sigma$ (km\,s$^{-1}$)  & \multicolumn{2}{c|}{0.30} & \multicolumn{2}{c|}{0.073}\\
\hline
\end{tabular}
\end{center}
\end{table}

Now that the $\gamma$ velocities at 17 timepoints are known, we could try to
determine the parameters of the outer orbit. Unfortunately, this is not
simple. While the new data show beyond any doubt that $\gamma$ is changing,
they alone do not allow to determine the orbit. A maximum $\gamma$ was
observed in 1996, but since then $\gamma$ has been decreasing steadily; no
minimum has been observed yet. That means that from the new data alone, only
lower limits for the period $P_{\rm out}$ of the outer orbit and the
RV-amplitude $K_{\rm out}$ can be obtained. The old datasets, due to their
unfortunate timings and their large errors do not improve this situation much.
This is with the exception of $\gamma_{\rm DHM_1}$ whose error is only
0.23\,km\,s$^{-1}$, and which will have a large impact on $P_{\rm out}$ and
by that also allow an estimate of $K_{\rm out}$. However, any orbital fit
to the $\gamma$ velocities as they are available now has to be considered
preliminary.

We have done a period search in the interval
$P_{\rm out}=4,..., 46$ years. Two runs were performed, one fixing $e=0$, the
other allowing for an elliptical orbit. The results of both searches are given
in Table~\ref{T-Fitout}; of all the orbit fits performed, the one yielding the
smallest $\sigma$ is given. While the elliptical orbit is {\it much}
better than the circular one, given the small number of points and their large
errors we cannot claim it to be correct.

The period of the circular orbit corresponds to \hbox{$(10.66\pm0.18)$\,yr}
and the
projected major axis $a_{\rm out}\,\sin i$ to \hbox{$(1.04\pm0.11)$\,AU};
the period
of the elliptical orbit is \hbox{$(21.46\pm0.06)$\,yr} and its projected major
axis is \hbox{$(1.15\pm0.06)$\,AU}.
%We performed a circular fit to the 17 $\gamma$ values vs.\ the average HJD as
%given in Table~\ref{T-Fitin}. Given the mostly low quality of the data and
%their small number it seems useless to allow for an eccentricity, although we
%note that it may improve the deviations of the new measurements from the
%orbit. The result is given in Table~\ref{T-Fitout} and shown in
%Fig.~\ref{F-Fitout}. The period corresponds to $(8.74\pm0.18)$\,yr; the
%projected major axis $a_{\rm out}\,\sin i$ to $(0.68\pm0.07)$\,AU.

\begin{figure}[t]
%\hspace*{-0.3cm}\rotatebox{90}{\resizebox{6.6cm}{!}{\includegraphics{uxari.rvcurve.long2.ps}}}
\hspace*{-0.3cm}\rotatebox{90}{\resizebox{6.6cm}{!}{\includegraphics{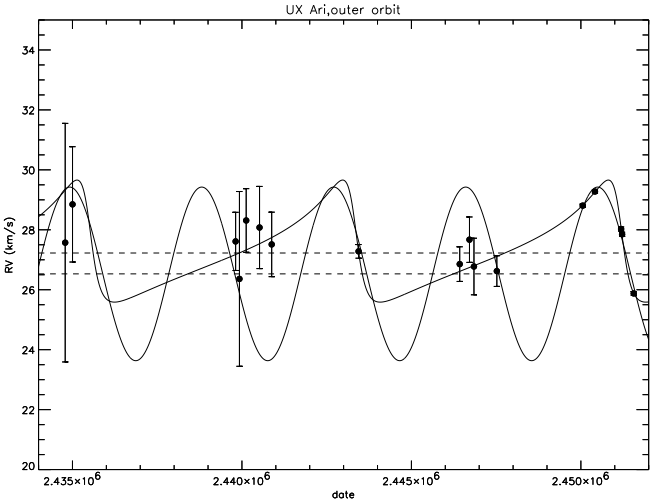}}}
\caption{\label{F-Fitout}
       The preliminary orbital fits to the 17 $\gamma$ velocities
       given in Table~\ref{T-Fitin}. For our new data, the last 5 points, the
       errorbars are smaller than the symbol sizes.}
\end{figure}

\subsection{The third star}
\label{SS-third star}

As already mentioned in Sect.~\ref{S-obs}, in the spectrum of UX~Ari\ there are
weak lines of a third star. Vogt \& Hatzes (\cite{VH91}) measured its RV
``stationary at 6.6\,km\,s$^{-1}$ blueward of the $\gamma$ velocity of the
system''. This statement refers to the time November 1986 to January 1987.
Assuming that they used the $\gamma$ velocity given by Carlos \& Popper
(\cite{CP71}), this velocity corresponds to 19.9\,km\,s$^{-1}$. We also measured the
velocities of the third star in the spectrum of UX~Ari\ after subtracting the
spectra of the primary and the secondary. We did not find any variation within
a run; thus, we give the average velocities and standard deviations
in Table~\ref{T-RV3}.

It is clear that the velocity of the third star is also changing. However,
there are several reasons to believe that the third star is not the body
responsible for the changes of the $\gamma$ velocity of the short-period
RS~CVn system:
\begin{itemize}
\item While the increase of RV$_3$ reflects the decrease of $\gamma$, $\gamma$
     reaches its maximum in 1996, while RV$_3$ reaches its minimum in (or
     before) 1995.
\item In 1986/7 and 1999, the third star had similar RVs. Either its $\gamma$
     is less than the 26.5 or 27.2\,km\,s$^{-1}$\ obtained from the long-period
     orbit of UX~Ari,
     or its period is at least 24 years (it must be longer, if 19.9 is below
     the $\gamma$ velocity of that binary), more than 2 times longer than our
     estimate for the circular long-period orbit of UX~Ari and still longer
     than the 21.5\,yr estimated for the elliptical orbit.
     However, we caution that it is
     always possible that a crossing of 19.9\,km\,s$^{-1}$\ between 1986/7 and
     1995 has been missed.
%\item The subtraction technique gives the flux ratios of the different
%     components, from which it follows that the third star is about 1\fm6 fainter
%%     than the secondary. If it has a similar spectral type as the secondary
%     (consistent with the lines we observe and also assumed by Vogt \& Hatzes
%      \cite{VH91}), the third star is at a distance twice that of the RS~CVn
%     system.
\item Yet another confirmation is obtained from the conclusions presented in
     Sect.~\ref{S-Concl}.
\end{itemize}

\begin{table}
\caption{\label{T-RV3} The radial velocities of the third star in the
       spectrum of UX~Ari. The velocities are weighted averages of
       all RVs obtained during the corresponding run.
       The velocity from Vogt \& Hatzes (\cite{VH91}, VH86/7), referring to
       the time Nov.\ 1986 to Jan.\ 1987, is
       also given; the velocity is inferred from their description and the
       assumption that they used the $\gamma$ velocity of Carlos \& Popper
       (\cite{CP71}); no error is available for their RV.
       }
\begin{center}
\begin{tabular}{|l|r@{$\pm$}l|}
\hline
dataset & \multicolumn{2}{c|}{RV$_3$ (\,km\,s$^{-1}$)}\\
\hline
VH86/7 & \multicolumn{2}{c|}{19.9}\\
S95 & 14.76 & 0.25\\
S96 & 15.532 & 0.086\\
M99 & 19.483 & 0.071\\
K99 & 19.917 & 0.093\\
M00 & 23.43 & 0.42\\
\hline
\end{tabular}
\end{center}
\end{table}

\section{Conclusions}
\label{S-Concl}

By using high-accuracy radial velocities of the double-lined RS~CVn system
UX~Ari\ we have shown that the $\gamma$ velocity is systematically changing;
all other orbital parameters seem to be constant over time. Preliminary
(circular and elliptical) orbital solutions of $\gamma(t)$ lead to a period
of roughly 10 and 21 years, respectively.
New, high-accuracy orbits of UX~Ari\ are urgently needed to
establish the accurate period of the long-period orbit.

It is interesting that our findings do not compare well with those
of Lestrade et al.\ (\cite{LPJ99}). Their angular accelerations measured for UX~Ari,
if interpreted as consequences of gravitational pull due to a third star, require
a period of the outer orbit of many times their 11-year observational time span.
Our circular orbit has a period even shorter than 11\,yr; the eccentric orbit's
period is less than twice 11\,yr. Furthermore, their observations
obtained during JD $\approx$ 2445000--2449000 covered, according to our eccentric
orbit fit, the passage through the apastron, i.e.\ they should have noticed a
change of direction in the proper motions. Our $\gamma$s are only compatible
with a much larger period if the eccentricity is even higher than $e\approx0.6$.
Our data time span is not %large
long
enough to allow for such a fit. Also for this
reason new, high-quality $\gamma$ measurements are of great importance.

Several arguments (see also below) indicate that the third star, which produces
a weak third set of lines in the spectrum of UX~Ari, is not the body responsible
for the $\gamma$ variations; but we found that the third star's RV is
also systematically varying. Thus, UX~Ari\ is (at least) a triple system with a
single-lined
spectroscopic binary on the same line of sight. Also for the third star further
observations are needed to establish its orbit and to find out whether it is
in the fore- or in the background.

If we take the preliminary long-period orbital solution seriously, the period
in the observer's frame is changing between $6\fd437778$ and $6\fd437902$
with an error of $0\fd000011$\footnote{These numbers refer to the larger $\gamma$
variations of the circular orbit (see Fig.~\ref{F-Fitout}).};
thus, the period variation is highly
significant (11$\sigma$). Nevertheless, the phase shift caused by this
variation is negligible for Doppler imaging: the apparent motion of a really
stationary surface feature during the 5.3\,yr between maximum and minimum is
only 2$^{\circ}$. The systematic change of the radial velocity by almost
6\,km\,s$^{-1}$,
however, is highly significant and may lead to artifacts in the maps.

In the following, we want to draw some more conclusions about the physical
parameters of the components in UX~Ari:
\begin{itemize}
\item The spectral type of the secondary is given as G5\,V (Vogt \& Hatzes
     \cite{VH91}), which is consistent with the results from our subtraction
     technique. If it were of significantly different spectral class, the
     subtraction of G5\,V standards would lead to significant residuals in
     the subtracted spectrum.
     The classification of G5\,V implies a mass of the secondary
     $m_2=(0.95\pm0.05)\,M_{\odot}$. This value is obtained as an average
     between the values given by Schmidt-Kaler (\cite{SK82}) and Gray
     (\cite{G92}, App.~B); the error accounts for the difference between their
     values as well as a small classification error.
\item Together with $m_2\,\sin^3 i$ from Table~\ref{T-Fitin} this yields
     an inclination of $i=59\fdg2\pm3\fdg3$. This high inclination
     (if the orbital and rotational axes are aligned) makes UX~Ari\ an ideal
     target for surface %(Doppler)
     imaging (for results see
     Aarum et al.\ \cite{Aa99},
     Aarum et al., in preparation).%~\cite{Aa00}).
\item $i$ and $m_1\,\sin^3 i$ (Table~\ref{T-Fitin}) give a
     mass of the primary of $m_1=(1.100\pm0.060)\,M_{\odot}$.
\item With $m_1$, $m_2$ and the mass functions $f(m)$ from the outer orbit
     (Table~\ref{T-Fitout}), and assuming that the inclination of the long-period
     orbit is also $59\fdg2$, we get for the mass of the
     third body in the system $m=(0.460\pm0.052)\,M_{\odot}$ (for $e=0$) and
     $m=(0.307\pm0.020)\,M_{\odot}$ (for $e\approx0.6$).
     The larger of these masses
     corresponds to an early M star, whose M$_{\rm V}$ is at least 4$^{\rm m}$
     fainter than that of the G5 secondary, much too faint to be
     responsible for the third set of lines visible in the spectrum of UX~Ari, thus
     confirming our earlier conclusion that the third star in the spectrum is
     a binary accidentally on the same line of sight.
\item The same conclusion is reached in another way: If the third star in the
     spectrum would be the body responsible for the $\gamma$ variations, the
     amplitude of its RV curve would be $K_3\geq26.5-14.8=11.7$\,km\,s$^{-1}$.
     This leads
     to a mass ratio of $m_3/(m_1+m_2)=K_{\rm out}/K_3 \leq 0.25$ and with the
     above masses $m_1$, $m_2$ to $m_3 \leq 0.51\,M_{\odot}$, again an early
     M star with too small luminosity to account for the third set of lines.
\item With the plausible assumption that the rotation of the two stars in the
     short-period system of UX~Ari\ is synchronized with the orbit,
     i.e.\ $P_{\rm rot}=P_{\rm orb}=P_{\rm rest}$ and $i_{\rm rot}=i_{\rm orb}$,
     $v\,\sin i$ gives for the radii of the stars
     $R_1=(5.78\pm0.13)\,R_{\odot}$,
     $R_2=(1.11\pm0.08)\,R_{\odot}$. The radius of the primary is
     consistent with its subgiant classification; the radius of the secondary,
     however, is significantly larger than that of a G5\,V star: Gray
     (\cite{G92}, App.~B) gives 0.96\,$R_{\odot}$, while Schmidt-Kaler
     (\cite{SK82}) gives 0.92\,$R_{\odot}$. With such a large radius the star
     should have spectral type G0--1\,V according to both authors. This seems
     to be a large deviation for the classification; maybe, also this
     star has already expanded a little away from the main sequence.
\item With the masses and radii we obtain for the surface gravities
      $\log g_1=2.96\pm0.03$, $\log g_2=4.32\pm0.07$. This means that the
     model spectra for the primary used in surface imaging can well be
     calculated with $\log g=3$; no interpolation between the
     models, given by Kurucz (\cite{K93}) in steps of $\Delta \log g=0.5$,
     will be necessary.
\item Given the inclination, we can compute $a=a_1+a_2$,
     and from that and the mass ratio (Table~\ref{T-Fitin}) obtain the
     effective radii of the Roche-lobes (Eggleton \cite{E83}),
     i.e.\ the radii of the spheres having the same volumes as the Roche-lobes.
     We get $R_{\rm Rl,1}=7.24\pm0.13$, $R_{\rm Rl,2}=6.77\pm0.12$
     (the errors neglect
     the (at most) 1\% error of the Eggleton (\cite{E83}) approximation).
\item If we compare the Roche-lobe radii with the radii of the stars themselves
     it is clear that neither of the stars fills its Roche-lobe. However, the
     primary is close; a K0 giant has typically a radius of 11\,$R_{\odot}$
     (Gray \cite{G92}) and since the star will go on cooling, the final radius
     should be even larger. Thus, the RS~CVn system of UX~Ari will soon become
     a semi-detached binary.
\end{itemize}

\begin{acknowledgements}
The authors would like to thank the anonymous referee for his careful reading
of the paper and his comments, which helped to significantly improve the paper.
This work made use of the SIMBAD database, maintained at the CDS, Strasbourg,
France. Part of this work is supported by the Norwegian Research Council under
project number 122520/431.
\end{acknowledgements}


\begin{thebibliography}{}
\bibitem[1999]{Aa99} Aarum V., Berdyugina S., Ilyin I. 1999, in Astrophysics
                   with the NOT, proc.\ conf., Turku, August 12--15, 1998,
                   (eds.: Karttunen H., Piirola V.), Univ.\ Turku, p.~222
%\bibitem[2000]{Aa00} Aarum V., Berdyugina S., Korhonen H., Duemmler R.,
%                   Ilyin I., Tuominen I., Engvold O. 2000, in preparation
\bibitem[1998]{B98} Berdyugina S.V., Berdyugin A.V., Ilyin I., Tuominen I.,
                  1998, A\&A 340, 437
\bibitem[1971]{CP71} Carlos R.C., Popper D.M., 1971, PASP 83, 504
\bibitem[1997]{DIT97} Duemmler R., Ilyin I.V., Tuominen I., 1997, A\&AS 123, 209
\bibitem[1991]{DMH91} Duquennoy A., Mayor M., Halbwachs J.-L. 1991, A\&AS 88,
                    281
\bibitem[1983]{E83} Eggleton P.P., 1983, ApJ 268, 368
\bibitem[1997]{ESA97} ESA 1997, The Hipparcos and Tycho catalogues, Vols.\
                      1--17, ESA--SP--1200
\bibitem[2000]{FM00} Fabricius C., Makarov V.V. 2000, A\&A 356, 141
\bibitem[1988]{G88} Gray D.F., 1988, Lectures on Spectral-Line Analysis:\ F, G
                    and K stars, The Publisher, Arva, Ontario
\bibitem[1992]{G92} Gray D.F., 1992, The Observation and Analysis of Stellar
                    Photospheres, Cambridge University Press
\bibitem[1973]{G73} Griffin R., Griffin R., 1973, MNRAS 162, 243
\bibitem[1997]{HMM97} Hartkopf W.I., McAlister H.A., Mason B.D., ten Brummelaar
                      T., Roberts L.C., Jr., Turner N.H., Wilson J.W. 1997, AJ
                      114, 1639
\bibitem[2000]{HMM00} Hartkopf W.I., Mason B.D., McAlister H.A., Roberts
                      L.C., Jr., Turner N.H., ten Brummelaar T.A., Prieto C.M.,
                      Ling J.F., Franz O.G. 2000, AJ 119, 3084
\bibitem[1981]{H81} Heintz W.D., 1981, ApJS 46, 247
\bibitem[1996]{I96} Ilyin I.V., 1996, Acquisition, Archiving and Analysis (3A)
                    Software Package -- User's Manual, Observatory, University
                    of Helsinki
\bibitem[2000]{I00} Ilyin I.V., 2000, High resolution SOFIN CCD \'echelle
                  spectroscopy, PhD thesis, University of Oulu, Finland
\bibitem[1993]{K93} Kurucz R.L. 1993, CD No.\ 13
\bibitem[1984]{KFBT84} Kurucz R.L., Furenlid I., Brault J., Testerman L. 1984,
                     Solar Flux Atlas from 296 to 1300\,nm, National Solar
                     Observatory Atlas No.\ 1
\bibitem[1999]{LPJ99} Lestrade J.-F., Preston R.A., Jones D.L., Phillips R.B.,
                      Rogers A.E.E., Titus M.A., Rioja M.J., Gabuzda D.C. 1999,
                      A\&A 344, 1014
\bibitem[1989]{L89} Lucy L.B., 1989, Obs.\ 109, No.\ 1090, 100
\bibitem[1971]{LS71} Lucy L.B., Sweeney M.A., 1971, AJ 76, 544
\bibitem[1987]{MHH87} McAlister H.A., Hartkopf W.I., Hutter D.J., 1987, AJ 93,
                      688
\bibitem[1973]{PB73} Pierce A.K., Breckinridge J.B., 1973, Kitt Peak Contr.\ 559
                   (and addendum 1974)
\bibitem[1982]{SK82} Schmidt-Kaler Th. 1982, in Landolt-B\"ornstein, Numerical
                   Data and Functional Relationships in Science and Technology,
                   New Series, Group VI: Astronomy, Astrophysics and Space
                   Research, Springer, Berlin, Vol.~2b, p.~30, Table~21
\bibitem[1993]{SHFS93} Strassmeier K.G., Hall D.S., Fekel F.C., Scheck M.,
                     1993, A\&AS 100, 173
\bibitem[1992]{T92} Tuominen I., 1992, NOT News No.\ 5, 15
\bibitem[1991]{VH91} Vogt S.S., Hatzes A.P., 1991, in: The Sun
                     and Cool Stars:\ activity, magnetism, dynamos, Proc.\ IAU
                     Coll.\ 130,
                     Tuominen I., Moss D., R\"udiger G. (eds.), Springer,
                     Berlin, p.~297
\end{thebibliography}
\end{document}